\documentclass[12pt,thmsa]{article}
\usepackage{amssymb}

\usepackage{sw20lart}

\input{tcilatex}
\begin{document}

\author{E. J. Ferrer}
\date{Dept. of Physics, State University of New York at Fredonia, Fredonia, NY
14063, USA}
\title{Non-Perturbative Effects of Strong Magnetic Fields on Neutrino Self-energy:
A New Approach\thanks{%
Talk given at the International Conference ''Quantization, Gauge Theory and
Strings'' dedicated to the memory of Prof. Efim Fradkin; Moscow, Jun. 5-10,
2000}}
\maketitle

\begin{abstract}
The Ritus' $E_{p}$ eigenfunction method is extended to the case of spin-1
charged particles in a constant electromagnetic field and used to calculate
the one-loop neutrino self-energy in the strong-field approximation $%
m_{e}^{2}\ll eB\ll M_{W}^{2}$. The implications of the obtained results for
neutrino oscillations in the early Universe, assuming the existence of a
sufficiently large ($m_{e}^{2}\ll eB\ll m_{\mu }^{2}$) primordial magnetic
field, are discussed.
\end{abstract}

The diagonalization, structure and properties of the Green's functions of
the electron and photon in an intense magnetic field were considered exactly
in external and radiative fields by Ritus in Refs. [1] and [2]. Ritus'
formulation provides an alternative method to the Schwinger approach to
address QFT problems on electromagnetic backgrounds. In Ritus' approach the
transformation to momentum space of the spin-1/2 particle Green's function
in the presence of a constant magnetic field is carried out using the $%
E_{p}(x)$ functions\cite{Ritus}$,$\cite{Ritus-Book} corresponding to the
eigenfunctions of the spin-1/2 charged particles in the electromagnetic
background. The $E_{p}(x)$ functions plays the role, in the presence of
magnetic fields, of the usual Fourier $e^{ipx}$ functions of the free case.
This method is very convenient for strong-field calculations, where the
Lower Landau Level (LLL) approximation is plausible and for finite
temperature calculations.

In this work we extend Ritus' method to the case of the spin-1 charged
particle. This extension permits to obtain a diagonal in momentum-space
Green's function for the spin-1 charged particle in the presence of a
constant magnetic field. Our results are important to investigate the
behavior of W-bosons in strong magnetic fields in the electroweak theory.

Starting from the Green's function equation for the W-bosons in the presence
of constant magnetic field \cite{Feldman}

\begin{equation}
\left[ \left( \Pi ^{2}+M_{W}^{2}\right) \delta _{\nu }^{\mu }-2ieF^{\mu
}\,_{\nu }+(\frac{1}{\alpha _{W}}-1)\Pi ^{\mu }\Pi _{\nu }\right] G_{\mu
}\,^{\nu }(x,y)=\delta ^{\left( 4\right) }(x,y)  \label{1}
\end{equation}
where $\Pi _{\mu }$ is given by

\begin{equation}
\Pi _{\mu }=-i\partial _{\mu }-eA_{\mu }^{ext},\qquad \mu =0,1,2,3  \label{2}
\end{equation}
with the potential in the Landau gauge

\begin{equation}
A_{\mu }^{\mathit{ext}}=Bx_{1}\delta _{\mu 2}  \label{2a}
\end{equation}
corresponding to a constant magnetic field of strength $B$ directed along
the $z$ direction in the rest frame of the system; we find that the
functions playing the role of the $E_{p}(x)$ functions for the spin-1/2 case
are\cite{Ferrer}

\begin{equation}
\Gamma _{k}^{\alpha }\,_{\mu }\left( x\right) =P^{\mu }\,_{\gamma
}F_{k}\left( x\right) P^{-1\gamma }\,_{\alpha }  \label{3}
\end{equation}
with

\begin{equation}
P^{\mu }\,_{\alpha }=\frac{1}{\sqrt{2}}\left( 
\begin{array}{llll}
\sqrt{2} & 0 & 0 & 0 \\ 
0 & 1 & 1 & 0 \\ 
0 & i & -i & 0 \\ 
0 & 0 & 0 & \sqrt{2}
\end{array}
\right) ,  \label{4}
\end{equation}

\begin{equation}
\left[ F_{k}\left( x\right) \right] _{\alpha }\,^{\beta }=\sum\limits_{\eta
=0,\pm 1}F_{k\eta }(x)\left[ \Omega ^{(\eta )}\right] _{\alpha }\,^{\beta },
\label{5}
\end{equation}

\begin{equation}
\Omega ^{(\eta )}=diag(\delta _{\eta ,0},\delta _{\eta ,1},\delta _{\eta
,-1},\delta _{\eta ,0}),\qquad \eta =0,\pm 1  \label{6}
\end{equation}
and

\begin{equation}
F_{k\eta }(x)=N(n)e^{i(p_{0}x^{0}+p_{2}x^{2}+p_{3}x^{3})}D_{n}(\xi )
\label{7}
\end{equation}
with $N(n)$ a normalization factor and $D_{n}(\xi )$ the parabolic cylinder
functions.

With the transformation functions (\ref{3}) the W-boson Green's function in
a constant magnetic field can be written as the following diagonal function
of momenta

\begin{equation}
G_{F}(k,k^{\prime })_{\mu }\,^{\nu }=%
%TCIMACRO{\TeXButton{sum_m}{\sum_m\hspace{-0.5cm}\int} }
%BeginExpansion
\sum_m\hspace{-0.5cm}\int%
%EndExpansion
\frac{d^{4}k}{\left( 2\pi \right) ^{4}}\Gamma _{k}^{\alpha }\,_{\mu }\left(
x\right) \frac{\delta _{\alpha }\,^{\beta }}{\overline{k}^{2}+M_{W}^{2}}%
\Gamma _{\,k\beta }^{\dagger }\,^{\nu }(y)=\left( 2\pi \right) ^{4}\widehat{%
\delta }^{(4)}(k-k^{\prime })\frac{\delta _{\mu }\,^{\nu }}{\overline{k}%
^{2}+M_{W}^{2}}  \label{40}
\end{equation}
The eigenvalue $\overline{k}^{2}$ is given by

\begin{equation}
\overline{k}^{2}=-k_{0}^{2}+k_{3}^{2}+2(\mathit{m}-1/2)eB,\qquad \mathit{m}%
=0,1,2,...  \label{41}
\end{equation}
where $\mathit{m}$ are the Landau numbers of the energy spectrum of the W
bosons in the presence of the magnetic field and $\widehat{\delta }%
^{(4)}(k-k^{\prime })\equiv \delta _{mm^{\prime }}\delta
(k_{0}-k_{0}^{\prime })\delta (k_{2}-k_{2}^{\prime })\delta
(k_{3}-k_{3}^{\prime })$.

It is known that to lowest order, the neutrino self-energy in a magnetic
field is given by the bubble diagram arising from the $e-W$ loop. Then, to
calculate the neutrino self-energy in the one-loop approximation we start
from the expression

\begin{equation}
\Sigma (x,y)=\frac{ig^{2}}{2}R\gamma _{\mu }S(x,y)\gamma ^{\nu
}G_{F}(x,y)_{\nu }\,^{\mu }L  \label{42}
\end{equation}
where $L,R=\frac{1}{2}(1\pm \gamma _{5})$, $G_{F}(x,y)_{\nu }\,^{\mu }$ is
the W-boson Green's function (\ref{40}), and $S(x,y)$ is the electron
Green's function in the presence of a constant magnetic field\cite{Ritus}

\begin{equation}
S(x,y)=%
%TCIMACRO{\TeXButton{sum_L}{\sum_{\it l}\hspace{-0.57cm}\int} }
%BeginExpansion
\sum_{\it l}\hspace{-0.57cm}\int%
%EndExpansion
\frac{d^{4}q}{\left( 2\pi \right) ^{4}}E_{q}\left( x\right) \frac{1}{\gamma .%
\overline{q}+m_{e}}\overline{E}_{q}(y)  \label{52}
\end{equation}
where

\begin{equation}
\overline{q}_{\mu }=\left( q_{0},0,-sgn\left( eB\right) \sqrt{2\left|
eB\right| l},q_{3}\right) ,\qquad l=0,1,2,...  \label{53}
\end{equation}
The integer $l$ in Eq. (\ref{53}) labels the electron Landau levels.

We will assume from now on that the magnetic field strength is in the range $%
m_{e}^{2}\ll eB\ll M_{W}^{2}$. This approximation for the magnetic field is
what we call the strong magnetic field approximation. Since in this case the
gap between the electron Landau-levels is larger than the electron mass
square ($eB\gg m_{e}^{2}$), we are allowed to use the LLL approximation for
the electron (i.e. $\mathit{l}=0$). On the other hand, it is obvious that
such approximation will not be valid for the W-bosons, so for them we are
bound to maintain the sum in all Landau levels. Thus, in this approximation
we have

\[
\left( 2\pi \right) ^{4}\delta ^{(4)}(p-p^{\prime })\Sigma (p)=\frac{-ig^{2}%
}{2}\int d^{4}xd^{4}y\int d^{4}q%
%TCIMACRO{\TeXButton{sum_m}{\sum_m\hspace{-0.5cm}\int}}
%BeginExpansion
\sum_m\hspace{-0.5cm}\int%
%EndExpansion
d^{4}k\frac{e^{-i(p.x-p^{\prime }.y)}}{(\overline{q}^{2}+m_{e}^{2})(%
\overline{k}^{2}+M_{W}^{2})} 
\]

\[
\left\{ \overline{q}_{\shortparallel }\cdot \gamma ^{\shortparallel }\left(
I_{\mathit{m}-2,\mathit{0}}(x)I_{\mathit{m}-2,\mathit{0}}^{*}(y)+I_{\mathit{m%
},\mathit{0}}(x)I_{\mathit{m},\mathit{0}}^{*}(y)\right) \Delta \left(
-\right) L\right. 
\]

\begin{equation}
+\left. \overline{q}_{\mu }\epsilon ^{1\mu 2\nu }\gamma _{\nu }\gamma
^{5}\left( I_{\mathit{m}-2,\mathit{0}}(x)I_{\mathit{m}-2,\mathit{0}%
}^{*}(y)-I_{\mathit{m},\mathit{0}}(x)I_{\mathit{m,0}}^{*}(y)\right) \Delta
\left( -\right) L\right\}  \label{54}
\end{equation}
In Eq. (\ref{54}) we are using the notation

\begin{equation}
I_{\mathit{a},\mathit{b}}(x)=\mathcal{F}_{\mathit{a}}\left( x\right) E_{%
\mathit{b}}(x)  \label{55}
\end{equation}
where $\mathcal{F}_{\mathit{a}}\left( x\right) =F_{k\eta }(x)$ and $E_{%
\mathit{b}}(x)=E_{p\sigma }(x)$. In (\ref{54}) the subindexes $\mathit{a}$
and $\mathit{b}$ were already written in terms of the Landau levels for the
W-bosons ($\mathit{m}$) and electrons ($\mathit{l}=0$). The matrix $\Delta
\left( -\right) $ is given by $\Delta (-)=diag(0,1,0,1)$.

After performing the integrations and sum in Eq. (\ref{54}) we end with the
following compact expression for the neutrino self-energy in the strong
field approximation

\begin{equation}
\Sigma (p)=\left[ ap\llap/_{\shortparallel }+bu\llap / +c\widehat{B}\llap / %
\right] L  \label{56}
\end{equation}
In (\ref{56}) the coefficients $a,$ $b$ and $c$ are Lorentz-invariant
functions which depend on the momentum and magnetic field. Their leading
contributions in powers of $1/M_{W}^{2}$ are given by

\begin{equation}
a=\frac{g^{2}}{(4\pi )^{2}}\lambda ,\qquad b=a\chi _{p},\qquad c=a\omega _{p}
\label{57}
\end{equation}
where

\begin{equation}
\lambda =\frac{\left| eB\right| }{M_{W}^{2}},\qquad \omega _{p}=p\cdot
u,\qquad \chi _{p}=p\cdot \widehat{B}  \label{58}
\end{equation}
The Lorentz scalars $\omega _{p}$ and $\chi _{p}$, as well as the structures
associated to the coefficients $b$ and $c$, depend on the four-velocity $%
u_{\mu }$ (where $u_{\mu }u^{\mu }=1$, and $u_{\mu }=(1,0,0,0)$ in the rest
frame ) and the unit vector along the external magnetic field $\widehat{%
\mathbf{B}}=\overrightarrow{B}/\left| \overrightarrow{B}\right| $,
respectively. The appearance of these two vectors is due to the fact that
the external magnetic field introduces a special Lorentz frame, the rest
frame where the applied field is defined, and a specialized direction, the
direction of the applied field. These two vectors ($u_{\mu }$ and $\widehat{%
\mathbf{B}}$) introduce additional structures in the neutrino self-energy as
compared to the case at $B=0$.

We stress that in the strong-field approximation, the parallel ($%
p_{\shortparallel }$), and perpendicular ($p_{\perp }$), momenta behave
quite differently. Thus, a strong magnetic field yields an anisotropy in the
neutrino propagation which leads to a neutrino self-energy mainly depending
on the momenta parallel to the applied field (\ref{56})-(\ref{58}). This
deviation from the vacuum mass shell condition $p_{\shortparallel
}^{2}=-p_{\perp }^{2}$ can have interesting consequences. For the case of
neutrino propagation in a magnetized medium, a self-energy structure similar
to (\ref{56}) has been reported \cite{Olivo}$.$ However, in that case \cite
{Olivo}$,$ the coefficients $a$ and $b$ are proportional to the electron and
positron densities which are functions of the electron chemical potential.

We would like to finish discussing some possible physical implications of
our results. We are particularly interested in their applications to
neutrino oscillations in the early Universe. The effect of external magnetic
fields on neutrino propagation has recently received increasing attention.
Its possible application to astrophysics, where fields of the order of $%
10^{13}$ $G$, and even larger \cite{Astro}$,$ can be expected in supernova
collapse and neutron stars, makes this subject worth of detailed
investigation. On the other hand, taking into account the recent
observations of large-scale magnetic fields in a number of galaxies, in
galactic halos, and in clusters of galaxies \cite{Galaxies}$,$ compelling
arguments have been given in favor of the existence of strong primordial
magnetic fields (i.e. fields of the order of $10^{24}$ $G$ at the
electroweak scale). Therefore, it is also of interest to study how such
strong fields could modify the neutrino propagation in the early Universe
and to investigate their possible influence in neutrino oscillations at
those epochs \cite{Osc}$.$

In the work we are reporting we calculate the zero temperature, zero density
($\mu =0$) contribution of the neutrino self-energy at strong magnetic
fields ($m_{e}^{2}\ll eB\ll M_{W}^{2}$, where $m_{e}$ is the electron mass
and $M_{W}$ is the W-boson mass). As discussed in ref. [4], such strong
fields can be expected to exist in the neutrino decoupling era.

The neutrino self-energy (\ref{56}), although independent of the mass of the
charged lepton, may contribute to the neutrino oscillation through the
following new mechanism: For flavor oscillations it is well known that any
variation in the dispersion relations corresponding to neutrinos with
different flavors is significative. For magnetic fields $m_{e}^{2}\ll eB\ll
m_{\mu }^{2}$, $M_{W}^{2}$ ($m_{\mu }$ is the muon mass), the muon-neutrino
self-energy, which corresponds to the weak-field approximation \cite{Olivo}, 
\cite{Feldman} ($eB\ll m_{\mu }^{2}$), is analytically different from the
one we are reporting for the electron-neutrino, which corresponds to the
strong-field approximation ($m_{e}^{2}\ll eB$). Hence, the corresponding
dispersion relations associated to these two different flavors (electron and
muon) will be significantly different, thus contributing to the oscillation
process.


\begin{thebibliography}{9}
\bibitem{Ritus}  V. I. Ritus, Ann. of Phys. (NY) \textbf{69} (1972) 555;
ZhETF \textbf{75} (1978) 1560 (Sov. Phys. JETP \textbf{48} (1978) 788);
ZhETF \textbf{76} (1979) 383.

\bibitem{Ritus-Book}  V. I. Ritus in Issues in Intense-Field Quantum
Electrodynamics, ed. V. L. Ginzburg (Nova Science, Commack, 1987).

\bibitem{Feldman}  A. Erdas and G. Feldman, Nucl. Phys. B\textbf{343} (1990)
597.

\bibitem{Ferrer}  E. Elizalde, E. J. Ferrer and V. de la Incera, ''Neutrino
Selfenergy and Index of Reraction in Strong Magnetic Field: A New
Approach,'' hep-ph/0007033.

\bibitem{Olivo}  J. C. D'Olivo, J. F. Nieves and P. B. Pal, Phys. Rev. D%
\textbf{40} (1989) 3679.

\bibitem{Astro}  V. L. Ginzburg, ``High Energy Gamma Ray Astrophysics''
(North Holland, Amsterdam, 1991); G. Ghanmugam, Ann. Rev. Astron. Astrophys. 
\textbf{30} (1992) 143.

\bibitem{Galaxies}  Y. Sofue, M. Fujimoto, and R. Wielebinski, Ann. Rev.
Astron. Astrophys. \textbf{24} (1986) 459; P. P. Kronberg, Rep. Prog. Phys., 
\textbf{57} (1994) 325; R. Beck et. al., Ann. Rev. Astron. Astrophys. 
\textbf{34} (1996) 153.

\bibitem{Osc}  K. Enqvist, K. Kainulainen and J. Maalampi, Phys. Lett. B%
\textbf{244} (1990) 186; P. Langacker and J. Liu, Phys. Rev. D\textbf{46}
(1992) 4140; V. B. Semikoz and J. W. F. Valle, Nucl. Phys. B\textbf{425}
(1994) 651; A. D. Dolgov, hep-ph/0004032.
\end{thebibliography}
\end{document}